\documentclass[12pt,psfig]{article}

\input epsf.tex

\usepackage{epsfig,subfigure,fancybox}

\usepackage{amsmath}
\usepackage{amssymb}
\usepackage{amsfonts}
\usepackage{graphicx}
\usepackage{dsfont}
\usepackage[subfigure]{graphfig}

\newtheorem{thm}{Theorem}

\newtheorem{lem}{Lemma}
\newtheorem{rem}{Remark}
\newtheorem{cor}{Corollary}

\newcommand{\thmref}[1]{Theorem~{\rm \ref{#1}}}
\newcommand{\lemref}[1]{Lemma~{\rm \ref{#1}}}

\newcounter{neweqn}

\newcommand{\beq}[1]{\begin{equation} \refstepcounter{neweqn} \label{#1}}
\newcommand{\eeq}{\end{equation}}
\newcommand{\bed}{\begin{displaymath}}
\newcommand{\eed}{\end{displaymath}}
\newcommand{\bedd}{\bed\begin{array}{l}}
\newcommand{\eedd}{\end{array}\eed}
\newcommand{\nd}{\noindent}

\newcommand{\disp}{\displaystyle}
\newcommand{\A}{{\cal A}}
\newcommand{\F}{{\cal F}}
\newcommand{\M}{{\cal M}}

\newcommand{\rr}{{\hbox{{\rm I}{\kern -0.22em}{\rm R}}}}
\newcommand{\bdd}{\hspace*{-0.08in}{\bf.}\hspace*{0.05in}}

\newcommand{\al}{\alpha}

\newcommand{\la}{\lambda}

\newcommand{\e}{\varepsilon}

\def\({\left(}
\def\){\right)}

\begin{document}

\title{When to Sell a Markov Chain Asset?\thanks{This research is supported in part by the Simons Foundation (235179).}}

\author{Q. Zhang\thanks{Department of Mathematics, University of Georgia,
Athens, GA 30602, qingz@math.uga.edu}}


\maketitle

\begin{abstract}

This paper is concerned with an optimal stock selling rule
under a Markov chain model. 
The objective is to find an optimal stopping time to sell the
stock so as to maximize an expected return.
Solutions to the associated variational inequalities are obtained.
Closed-form solutions are given in terms of a set of threshold levels.
Verification theorems are provided to justify their optimality. 
Finally, numerical examples are reported to illustrate the results.


\end{abstract}


\newpage

\baselineskip25pt
\renewcommand{\arraystretch}{1.8}

\section{Introduction}

Most market models in the literature are Brownian motion based including
geometric Brownian motion, diffusion with possible jumps and regime switching;
see related books by Duffie \cite{Duffie}, Hull \cite{Hull},
Elliott and Kopp \cite{ElliottK}, Fouque et al. \cite{FouquePR},
Karatzas and Shreve \cite{KaratzasS},
and Musiela and Rutkowski \cite{MusielaR} among others.
An alternative is the binomial tree model introduced by
Cox-Ross-Rubinstein.
The BTM is natural for financial markets because 
intensive buying moves the market upwards and forceful selling 
pushes it downwards. 
All these transactions take place in discrete moments. 
However, a main drawback of the BTM is its non-Markovian nature, 
which makes it difficult to work with mathematically. 
In this paper, we consider a Markov chain market model.
The main advantage of such a model is it preserves much of the flexibility of
the binomial tree structure and, in the meantime, 
it is more mathematically tractable, which allows serious mathematical
analysis in related optimization problems. 
Recently, several Markov chain based models are developed. 
For example, van der Hoek and Elliott \cite{HoekE} introduced 
a stock price model based on stock dividend rates and a Markov chain noise.
Norberg \cite{Norberg} used a Markov chain to represent interest rate and
considered a market model driven by a Markov chain. In particular,
the market model in \cite{Norberg} resembles a GBM in which
the `drift' is approximated by the duration between jumps and the
`diffusion' is given in terms of jump times.
An additional advantage of a Markov chain driven model is its price is
almost everywhere differentiable. Such differentiability is 
desirable in an optimal control type 
analysis proposed by Barmish and Primbs \cite{BarmishP}.
In connection with dynamic programming problems, 
the corresponding Hamilton-Jacobi-Bellman equations are of first order,
which are easier to analyze than those under 
traditional Brownian motion based models.
Finally, the Markov chain model is not that far apart from 
a GBM because it
can be used to approximate a GBM by varying its jump rates. 
In fact, it is shown in 
Example 1 that a properly scaled Markov chain model converges weakly to
that of a GBM as the jump rates go to infinity.

When to sell a stock is a crucial component in stock trading.
It determines when to take profits or to cut losses. 
It is probably the most emotional part for individual investors
in the trading process.
Selling rules in financial markets have been studied for many
years. For example, Zhang \cite{Zhang-trading} considered a
selling rule determined by two threshold levels: a target price
and a stop-loss limit. One makes a selling decision whenever the
price reaches either levels. Under a switching GBM,
the objective is to determine these threshold levels to maximize
an expected discounted reward function. In \cite{Zhang-trading},
such optimal threshold levels are obtained by solving a set of
two-point boundary value problems. In Guo and Zhang \cite{GuoZ},
they considered the optimal selling rule under a GBM model with
regime switching. Using a smooth-fit technique, they were able to
convert the optimal stopping problem to a set of algebraic
equations.
These algebraic equations were used to determine the optimal target levels.
In addition to these analytical results, various mathematical tools
have been developed to compute these threshold levels. For example,
a stochastic approximation technique was used in Yin, Liu and Zhang
\cite{YLZ} and a linear programming approach was developed in Helmes
\cite{Helmes}.
In addition, Merhi and Zervos \cite{MerhiZ} studied 
an investment capacity expansion/reduction problem following
a dynamic programming approach under a GBM market model.
Similar problem under a more general market model was treated by 
L{\o }kka and Zervos \cite{LokkaZ}.

In this paper, the stock price is assumed to follow
a Markov chain model. Under this model,
the state of the Markov chain can be estimated based on the
stock price increments. This makes the Markov chain observable.
In addition to its simplicity, the Markov chain model 
is able to capture price movements of a broader range of stocks.
In this paper, under the Markov chain model, 
we consider an optimal stock selling rule and obtain its
solution in terms of a set of threshold levels. 
In particular, we solve the corresponding 
dynamic programming problem and obtain
these threshold levels.
We point out that the standard smooth-fit method that works in
a GBM setting is not adequate in one of the 
cases in this paper because of the lack
of enough equations for the unknown parameters.
To solve the problem, we need to explore other convexity
conditions to determine uniquely these parameters.
We also provide a set of sufficient conditions that guarantee their optimality.
Numerical examples are reported to illustrate these results.

This paper is organized as follows.
In \S2, we formulate the problem and make a few assumptions.
In \S3, we study properties of the value functions,
the associate HJB equations, and their solutions.
In \S4, we provide a set of sufficient conditions
that guarantee the optimality of our selling rule.
We also include three numerical examples in this section. 
Some concluding remarks are given in \S5. 
Some technical results are provided in an appendix.

\section{Problem Formulation}

Let $\{\al_t:\ t\geq0\}$ denote a two-state Markov chain with
state space $\M=\{1,2\}$ and generator 
$Q=\(\begin{array}{cc}
   -\la_1&\la_1\\[-0.1in]
\la_2&-\la_2\\ 
\end{array}\)$, for given $\la_1>0$ and $\la_2>0$.             
Let $S_t$ denote the stock price at time $t$ given by the 
equation
\[
\frac{dS_t}{S_t}=f(\al_t)dt,\ S_0=x\geq0,\ t\geq0,
\]
where $f(1)=f_1>0$ represents the uptick return rate and
$f(2)=f_2<0$ the downtick return rate.
Let $\F_t=\{S_r:\ r\leq t\}$ denote the filtration generated by 
$S_t$. Note that $\al_t$ is observable and 
$\F_t=\{\al_r:\ r\leq t\}$.

Let $K$ denote the fixed transaction cost.
Given $S_0=x$ and $\al_0=i\in\M$,
the objective of the problem is to choose an $\F_t$ stopping time
$\tau$ so as to maximize 
\[
J(x,i,\tau)=E\(e^{-\rho\tau}(S_\tau-K)I_{\{\tau<\infty\}}\),
\]
where $\rho>0$ is the discount factor.

Let $V(x,i)=\sup_\tau J(x,i,\tau)$ be the value function. 
Then it is easy to see that $V(x,i)\geq0$, $V(0,i)=0$, $i=1,2$.
Moreover, $V(x,i)$ is convex in $x$ for fixed $i=1,2$.

Let $\Phi(\rho)=(\rho+\la_1-f_1)(\rho+\la_2-f_2)-\la_1\la_2$.
Then the bigger root of $\Phi(\rho)=0$ is given by
\[
B_0=\frac{1}{2}\(f_1-\la_1+f_2-\la_2+
\sqrt{((f_1-\la_1)-(f_2-\la_2))^2+4\la_1\la_2}\).
\]

Note that if $\rho\leq B_0$, then following similar argument as in
Guo and Zhang \cite{GuoZ}, we can show that it is optimal not to 
sell at all.
In the rest of this paper, we only consider the case when
$\rho>B_0$, which implies $\Phi(\rho)>0$.
We summarize the conditions to be imposed in the rest of this paper:

\begin{itemize}
\item[ ]{\bf (A1)} $f_1>0$ and $f_2<0$;
\item[ ]{\bf (A2)} $\Phi(\rho)>0$. 
\end{itemize}

\medskip
Let 
$\disp (\nu_1,\nu_2)=\({\la_2}/{(\la_1+\la_2)},{\la_1}/{(\la_1+\la_2)}\)$
denote the stationary distribution of $\al_t$ and 
let $\mu=\nu_1 f_1+\nu_2f_2$. Then, $f_2<\mu<f_1$.
Moreover, it is easy to see that
$\Phi(\mu)=(\mu-f_1)(\mu-f_2)<0$. This implies $B_0>\mu$.
Therefore, $\rho>\mu$.

Note that, for any ${\cal F}_t$ stopping time $\tau$,
\[
J(x,i,\tau)=xE\(e^{-\rho \tau}\exp\int_0^\tau f(\al_s)ds\)I_{\{\tau<\infty\}}
-KE e^{-\rho\tau}.
\]
In order to have finite $V(x,i)$, necessarily
\[
\sup_{\tau}E\(e^{-\rho \tau}\exp\int_0^\tau f(\al_s)ds\)I_{\{\tau<\infty\}}
<\infty.
\]
In view of this, $V(x,i)$ needs to be at most linear growth in $x$.
In addition, note that the stock price $S_t$ is differentiable and 
the value of $\al_t$ can be given in terms of 
the derivative of $\log(S_t)$.

\section{HJB Equations}
Let $\A$ denote the generator of $(S_t,\al_t)$, i.e., for any differentiable 
functions $h(x,i)$, $i=1,2$,
\[
\left\{\begin{array}{l}
\disp
\A h(x,1)=xf_1h'(x,1)+\la_1(h(x,2)-h(x,1)),\\[0.1in]
\disp
\A h(x,2)=xf_2h'(x,2)+\la_2(h(x,1)-h(x,2)),\\
\end{array}\right.
\]
where $h'$ denotes the derivative of $h$ with respect to $x$.
The associated HJB equations should have the form:
\beq{HJB}
\left\{\begin{array}{l}
\min\{\rho v(x,1)-\A v(x,1),\ v(x,1)-(x-K)\}=0,\\
\min\{\rho v(x,2)-\A v(x,2),\ v(x,2)-(x-K)\}=0.
\end{array}\right.
\eeq

In this section, we solve these HJB equations.
First, if the price $S_t$ is small, then one should hold the
position because the price is not attractive regardless $\al_t=1$ or 2. 
In view of this, we expect the existence of $x^*$ such that 
no selling is $S_t<x^*$. The corresponding interval $(0,x^*)$ gives 
a continuation region.
Note that $V(x,i)\geq0$ implies $x^*\geq K$. 
On this interval, the equalities
$\rho v(x,i)-\A v(x,i)=0$, $i=1,2$, must hold.
Using the generator $\A$, we can write
\[
\left\{\begin{array}{l}
(\rho+\la_1)v(x,1)=xf_1 v'(x,1)+\la_1 v(x,2),\\
(\rho+\la_2)v(x,2)=xf_2 v'(x,2)+\la_2 v(x,1).\\
\end{array}\right.
\]
Using the first equation, we write 
\[
v(x,2)=\frac{1}{\la_1}\((\rho+\la_1)v(x,1)-xf_1 v'(x,1)\).
\]
Substitute this into the second equation and simplify to obtain
\beq{joint-ode}
x^2f_1f_2v''(x,1)+x(f_1f_2-D_1)v'(x,1)+D_2 v(x,1)=0,
\eeq
where 
\beq{D1-D2}
\left\{\begin{array}{l}
D_1=(\rho+\la_1)f_2+(\rho+\la_2)f_1,\\ 
D_2=(\rho+\la_1)(\rho+\la_2)-\la_1\la_2.
\end{array}\right.
\eeq

Let $\beta_1<0$ and $\beta_2>0$ denote the roots of 
\beq{char-eqn}
f_1f_2\beta^2-D_1\beta+D_2=0.
\eeq
Then, 
\beq{char-root}
\left\{\begin{array}{l}
\disp
\beta_1=\frac{D_1+\sqrt{D_1^2-4f_1f_2 D_2}}{2f_1f_2}<0,\\ [0.1in]
\disp
\beta_2=\frac{D_1-\sqrt{D_1^2-4f_1f_2 D_2}}{2f_1f_2}>0.\\ 
\end{array}\right.
\eeq
The general solution to (\ref{joint-ode}) can be given as
\[
v(x,1)=A_1 x^{\beta_1}+A_2 x^{\beta_2},
\]
for some constants $A_1$ and $A_2$.

On $(0,x^*)$, the convexity condition implies that $v(x,1)$ 
is bounded. Necessarily, $A_1=0$. Therefore, $v(x,1)=A_2x^{\beta_2}$.
Substitute this back into the first equation to obtain
$v(x,2)=\kappa_2 A_2 x^{\beta_2}$, where
$\kappa_2={(\rho+\la_1-f_1\beta_2)}/{\la_1}$.

Recall that $\al_t$ is observable. One should hold the position longer
under the condition $\al_t=1$ (uptick) than that under $\al_t=2$ (downtick). 
In view of this, 
we consider the HJB equations on $(x^*,x_0^*)$ for some $x_0^*>x^*$.
The idea is to sell if $(S_t,\al_t)\in [x^*,\infty)\times\{2\}$
and hold if $(S_t,\al_t)\in(0,x_0^*)\times\{1\}$
till $S_t$ reaching $x_0^*$.
Clearly, $v(x,2)=x-K$ and 
$\rho v(x,1)-\A v(x,1)=0$, on $(x^*,x_0^*)$. 
Using this, we solve the equation 
\[
\rho v(x,1)-\A v(x,1)=0,\\
\]
which gives
\[
(\rho+\la_1)v(x,1)=xf_1 v'(x,1)+\la_1(x-K).
\]
It is easy to see a particular solution
\beq{A0-B0}
\phi_0(x)=A_0x+B_0,\mbox{ where }A_0=\frac{\la_1}{\rho+\la_1-f_1}
\mbox{ and } B_0=-\frac{\la_1 K}{\rho+\la_1}.
\eeq
Let $\gamma_1=(\rho+\la_1)/f_1$. Then, the general solution can be
given by
\[
v(x,1)=C_1x^{\gamma1}+\phi_0(x),
\]
for any constant $C_1$.

Next we consider two separate cases to continue solving the HJB equations.

\subsection*{Case I: $\rho\leq f_1$}
Assuming $\rho\leq f_1$, we first show that $x_0^*=\infty$. 
If not, we must have $v(x,1)=v(x,2)=x-K$ for $x>x_0^*$.
In order to satisfy the HJB equations (\ref{HJB}), $v(x,1)$ has to 
satisfy the inequality
$\rho v(x,1)-\A v(x,1)\geq0$, for $x>x_0^*$. Plugging 
$v(x,1)=v(x,2)=x-K$ in this inequality, we have 
\[
\rho (x-K)\geq x f_1.
\]
Therefore, $(\rho-f_1)x\geq \rho K$, for $x>x_0^*$. This
contradicts $\rho\leq f_1$. Hence $x_0^*=\infty$.

In view of these, on $(x^*,\infty)$, $v(x,1)=C_1x^{\gamma1}+\phi_0(x)$
and $v(x,2)=x-K$. This means never sell when $\al_t=1$.
Recall the linear growth property and nonnegativity
of $v(x,i)$. It follows that $C_1=0$ because $\gamma_1>1$.

Next, we determine the values of $x^*$ and $A_2$. 
Recall that $v(x,1)$ and $v(x,2)$ are convex on $(0,\infty)$.
Necessarily, they are continuous. In particular, they 
are continuous at $x=x^*$. Therefore,
\[
\left\{\begin{array}{l}
A_2(x^*)^{\beta_2}=A_0x^*+B_0,\\
\kappa_2A_2(x^*)^{\beta_2}=x^*-K.
\end{array}\right.
\]
Solving these equations, we have
\beq{x^*-Case-I}
x^*=-\frac{K+\kappa_2 B_0}{\kappa_2 A_0-1},
\eeq
and 
\beq{A2}
A_2=\frac{A_0x^*+B_0}{(x^*)^{\beta_2}}.
\eeq

It is elementary to check that
\beq{x*-alternative}
x^*=\(\frac{\rho+\la_1-f_1}{\rho+\la_1}\)\(\frac{K\beta_2}{\beta_2-1}\).
\eeq

The solutions to the HJB equations (\ref{HJB}) should have the form:
\beq{value-fn-Case-I}
\left\{\begin{array}{l}
v(x,1)=\left\{\begin{array}{ll}
A_2 x^{\beta_2}&\mbox{ if }0\leq x\leq x^*,\\
A_0x+B_0&\mbox{ if }x> x^*,\\
\end{array}\right.\\
v(x,2)=\left\{\begin{array}{ll}
\kappa_2A_2 x^{\beta_2}&\mbox{ if }0\leq x\leq x^*,\\
x-K&\mbox{ if }x> x^*.\\
\end{array}\right.\\
\end{array}\right.
\eeq

\begin{thm}\bdd\label{HJB-Case-I}
Assume $\rho\leq f_1$.
Then the functions $v(x,i)$, $i=1,2$, given above are continuous on 
$(0,\infty)$ and differentiable on $(0,\infty)-\{x^*\}$.
They satisfy the HJB equations {\rm (\ref{HJB})}.
In particular, the following inequalities hold:
\beq{v-ineq}
\left\{\begin{array}{ll}
A_2 x^{\beta_2}\geq x-K&\mbox{ on }(0,x^*),\\
\kappa_2A_2 x^{\beta_2}\geq x-K&\mbox{ on }(0,x^*),\\
v(x,1)\geq x-K&\mbox{ on }(x^*,\infty),\\
\rho v(x,2)-\A v(x,2)\geq 0&\mbox{ on }(x^*,\infty).
\end{array}\right.
\eeq
\end{thm}

\nd{\it Proof.}
It is sufficient to show these four inequalities.
First, note that $\Phi(\rho)>0$ implies $\rho+\la_1-f_1>0$. 
Under the condition $\rho\leq f_1$, we have $A_0\geq1$.
The third inequality in (\ref{v-ineq}) follows from $B_0>-K$.
In addition, the first inequality follows from the second one
because $0<\kappa_2<1$ as shown in Appendix (\lemref{kappa2}).
To show the second inequality, we claim that $A_2>0$ and 
\beq{convex-2}
\beta_2\kappa_2A_2(x^*)^{\beta_2-1}<1.
\eeq
To see $A_2>0$, notice that (\ref{x*-alternative}) implies
\[
x^*>\frac{(\rho+\la_1-f_1)K}{\rho+\la_1},
\]
because $\beta_2>1$ (see \lemref{beta2>1} in Appendix). 
Therefore $A_0 x^*+B_0$ is positive, so is $A_2$.
To show (\ref{convex-2}), use again (\ref{x*-alternative}), which yields
\[
x^*<\frac{K\beta_2}{\beta_2-1}.
\]
This is equivalent to (\ref{convex-2}) because
$\kappa_2 A_2(x^*)^{\beta_2}=x^*-K$.
Let $\phi(x)=\kappa_2A_2x^{\beta_2}-(x-K)$.
In view of the above claim and the definition of $x^*$, it follows
that, on $(0,x^*)$,
\[
\begin{array}{l}
\phi(x^*)=\kappa_2A_2(x^*)^{\beta_2}-(x^*-K)=0,\\
\phi'(x^*)=\beta_2\kappa_2A_2(x^*)^{\beta_2-1}-1< 0,\\
\phi''(x)=\beta_2(\beta_2-1)\kappa_2A_2x^{\beta_2-2}>0.
\end{array}
\]
Consequently, $\phi'(x)$ is increasing on $(0,x^*)$, which implies
$\phi'(x)<0$. Hence, $\phi(x)$ is decreasing. Therefore,
$\phi(x)>0$ on $(0,x^*)$, which implies the second inequality in 
(\ref{v-ineq}).

It remains to show the last inequality $\rho v(x,2)-\A v(x,2)\geq0$  
in (\ref{v-ineq}).
This is equivalent to 
\[
\rho(x-K)\geq xf_2+\la_2(\phi_0(x)-(x-K)).
\]
It follows that
\[
(\rho+\la_2-f_2-\la_2 A_0)x\geq (\rho+\la_2)K+\la_2B_0.
\]
Using the notation $\Phi(\rho)$ and $D_2$, we have
\[
\(\frac{\Phi(\rho)}{\rho+\la_1-f_1}\)x\geq \frac{KD_2}{\rho+\la_1}.
\]
Therefore, we need 
\[
x\geq\frac{K(\rho+\la_1-f_1)D_2}{(\rho+\la_1)\Phi(\rho)},
\]
for $x\geq x^*$.
It suffices to show this inequality when $x=x^*$. Using the expression in
(\ref{x*-alternative}), we only have to show
\[
\frac{\beta_2}{\beta_2-1}\geq\frac{D_2}{\Phi(\rho)}.
\]
Rewrite this to obtain
\beq{1-beta2}
\frac{1}{\beta_2}\geq 1-\frac{\Phi(\rho)}{D_2}.
\eeq
Now, if $1-{\Phi(\rho)}/{D_2}\leq 0$ (i.e., $D_2\leq \Phi(\rho)$), then 
we are done because $\beta_2>1$. Otherwise, $D_2>\Phi(\rho)$. 
Under this condition, we rewrite (\ref{1-beta2}) as
\[
\beta_2\leq\frac{D_2}{D_2-\Phi(\rho)},
\]
which is equivalent to
\beq{rho-v2}
\sqrt{D_1^2-4f_1f_2D_2}\leq D_1-\frac{2f_1f_2}{D_2-\Phi(\rho)}.
\eeq
Note that
\beq{Phi-D1-D2}
\Phi(\rho)=D_2-D_1+f_1f_2.
\eeq
Therefore, $D_2>\Phi(\rho)$ implies that $D_1>f_1f_2$.
Under this condition, it is easy to check 
\[
D_1-\frac{2f_1f_2D_2}{D_2-\Phi(\rho)}>
f_1f_2-\frac{2f_1f_2D_2}{D_2-\Phi(\rho)}=
-f_1f_2\(\frac{D_2+\Phi(\rho)}{D_2-\Phi(\rho)}\)>0.
\]
Square both sides of (\ref{rho-v2}) to obtain
\[
D_1^2-4f_1f_2D_2\leq D_1^2-\frac{4f_1f_2D_1D_2}{D_2-\Phi(\rho)}
+\frac{4f_1^2f_2^2D_2^2}{(D_2-\Phi(\rho))^2}.
\]
Simplify this inequality to have
\[
D_1(D_2-\Phi(\rho))-f_1f_2D_2\geq(D_2-\Phi(\rho))^2.
\]
Furthermore, using (\ref{Phi-D1-D2}), we have $D_2-\Phi(\rho)=D_1-f_1f_2$.
Substitute this into the above inequality to obtain
\[
D_1(D_1-f_1f_2)-f_1f_2D_2\geq (D_1-f_1f_2)^2.
\]
This is equivalent to 
\[
D_2-D_1\geq -f_1f_2,
\]
which leads $\Phi(\rho)\geq 0$, which holds under the
assumption $\Phi(\rho)>0$.
Therefore, $\rho v(x,2)-\A v(x,2)\geq0$ on $(x^*,\infty)$.
The proof is compete. \hfill$\Box$

\begin{rem}\bdd
{\rm
Using (\ref{x*-alternative}), 
one can show $\beta_2A_2(x^*)^{\beta_2-1}=A_0$. This implies that
$v(x,1)$ is differentiable at $x=x^*$. On the other hand,
following (\ref{convex-2}), we can see that 
$v(x,2)$ is not differentiable at $x=x^*$. 
}
\end{rem}

\subsection*{Case II: $\rho> f_1$} 
We consider the second case when $\rho> f_1$. Note that a large $\rho$
encourages selling sooner. Naturally, we expect $x_0^*<\infty$. 
The solutions to the HJB equations (\ref{HJB}) should have the form:

\beq{value-fn-Case-II}
\left\{\begin{array}{l}
v(x,1)=\left\{\begin{array}{ll}
A_2 x^{\beta_2}&\mbox{ if }0\leq x\leq x^*,\\
C_1x^{\gamma_1}+\phi_0(x)&\mbox{ if }x^*< x\leq x_0^*,\\
x-K&\mbox{ if }x> x_0^*,\\
\end{array}\right.\\
v(x,2)=\left\{\begin{array}{ll}
\kappa_2A_2 x^{\beta_2}&\mbox{ if }0\leq x\leq x^*,\\
x-K&\mbox{ if }x> x^*.\\
\end{array}\right.\\
\end{array}\right.
\eeq
We need to determine the values of $A_2$, $C_1$, $x_0^*$, and $x^*$.
Again, following the continuity of the value functions
at $x^*$ and $x_0^*$, we have
\beq{cont-conds-case-II}
\left\{\begin{array}{l}
A_2(x^*)^{\beta_2}=C_1(x^*)^{\gamma_1}+\phi_0(x^*),\\
\kappa_2A_2(x^*)^{\beta_2}=x^*-K,\\
C_1(x_0^*)^{\gamma_1}+\phi_0(x_0^*)=x_0^*-K.\\
\end{array}\right.
\eeq
Note that there are only three equations, which are not adequate to
determine uniquely the values of the four unknowns.
We need to find further conditions.
Note that
to satisfy the HJB equations (\ref{HJB}), the following inequalities
have to hold:

\beq{VI-1-Case-II}
\left\{\begin{array}{l}
A_2x^{\beta_2}\geq x-K,\\
\kappa_2A_2x^{\beta_2}\geq x-K,\\
\end{array}\right.
\mbox{    on }(0,x^*);
\eeq

\beq{VI-2-Case-II}
\left\{\begin{array}{l}
C_1x^{\gamma_1}+\phi_0(x)\geq x-K,\\
\rho v(x,2)-\A v(x,2)\geq0,\\
\end{array}\right.
\mbox{    on }(x^*,x_0^*);
\eeq

\beq{VI-3-Case-II}
\left\{\begin{array}{l}
\rho v(x,1)-\A v(x,1)\geq0,\\
\rho v(x,2)-\A v(x,2)\geq0,\\
\end{array}\right.
\mbox{    on }(x_0^*,\infty).
\eeq

First, we consider (\ref{VI-1-Case-II}).
Note that convexity of $v(x,2)$ at $x=x^*$ implies
\beq{case-II-x^*}
\beta_2\kappa_2A_2(x^*)^{\beta_2-1}\leq 1.
\eeq
Under this condition, following from similar 
argument used to prove the second inequality in 
(\ref{v-ineq}) with $\phi(x)=\kappa_2A_2x^{\beta_2}-(x-K)$
for possibly different $x^*$, we can show
the second inequality in (\ref{VI-1-Case-II}) holds, so does the first one.
Therefore, the inequalities in (\ref{VI-1-Case-II}) are equivalent to
(\ref{case-II-x^*}), which can be simplified and written as:
\beq{case-II-x^*-v2}
x^*\leq \frac{K\beta_2}{\beta_2-1}.
\eeq

Next, we consider (\ref{VI-3-Case-II}). The first inequality implies
\[
\rho(x-K)\geq xf_1, \mbox{ for }x< x_0^*.
\]
It follows that
\beq{x_0^*-cond}
x_0^*\geq \frac{\rho K}{\rho-f_1}.
\eeq
The second inequality in (\ref{VI-3-Case-II})
is automatically satisfied because $f_2<0$.

Finally, go back to (\ref{VI-2-Case-II}).
Again, the convexity of $v(x,1)$ at $x=x_0^*$ yields 
\beq{1-ineq-VI-2-Case-II}
\gamma_1 C_1(x_0^*)^{\gamma_1-1}+A_0\leq 1.
\eeq

It follows from the third equality in (\ref{cont-conds-case-II}) that
\beq{C1}
C_1=\frac{x_0^*-K-\phi_0(x_0^*)}{(x_0^*)^{\gamma_1}}.
\eeq
Under the condition $x_0^*\geq \rho K/(\rho-f_1)$, it is easy to see
$C_1>0$.
Note that $0<A_0<1$ under $\rho>f_1$.
Let $\phi(x)=C_1x^{\gamma_1}+\phi_0(x)-(x-K)$. Then,
it is direct to check that $\phi(x_0^*)=0$, $\phi'(x_0^*)\leq0$,
$\phi''(x)=\gamma_1(\gamma_1-1)C_1x^{\gamma_1-2}>0$. 
Therefore, $\phi'(x)$ is increasing on $(x^*,x_0^*)$. 
Thus, $\phi'(x)<0$ on $(x^*,x_0^*)$, which implies
$\phi(x)$ is decreasing. Therefore, $\phi(x)\geq0$ on $(x^*,x_0^*)$. 
The first inequality in (\ref{VI-2-Case-II}) follows from
(\ref{1-ineq-VI-2-Case-II}).

Use (\ref{C1}) and rewrite (\ref{1-ineq-VI-2-Case-II}) to obtain
\[
\gamma_1(x_0^*-K-\phi_0(x_0^*))\leq (1-A_0)x_0^*,
\]
which leads to 
\[
(\gamma_1-1)(1-A_0)x_0^*\leq \gamma_1(K+B_0).
\]
Recall that $\gamma_1>1$ and $A_0<1$. It follows that
\[
x_0^*\leq \frac{\gamma_1(K+B_0)}{(\gamma_1-1)(1-A_0)}=\frac{\rho K}{\rho-f_1}.
\]
Combining the opposite inequality (\ref{x_0^*-cond}), we have
\[
x_0^*=\frac{\rho K}{\rho-f_1}.
\]

Next, we claim that the second inequality in (\ref{VI-2-Case-II}) 
follows from 
\beq{2nd-ineq}
C_1(x^*)^{\gamma_1}+\phi_0(x^*)\leq \frac{(\rho+\la_2)(x^*-K)-x^* f_2}{\la_2}.
\eeq
To see this, let 
\[
\phi(x)=C_1x^{\gamma_1}+\phi_0(x)-\frac{(\rho+\la_2)(x-K)-x f_2}{\la_2}.
\]
Then, using $v(x,1)=C_1x^{\gamma_1}+\phi_0(x)$ and $v(x,2)=x-K$,
the second inequality becomes $\phi(x)\leq0$
on $x\in(x^*,x_0^*)$.
Under (\ref{2nd-ineq}), $\phi(x^*)\leq0$. Note also that
\[
\phi'(x_0^*)=\gamma_1C_1(x_0^*)^{\gamma_1}+A_0-\frac{\rho+\la_2-f_2}{\la_2}
\leq 1-\frac{\rho+\la_2-f_2}{\la_2}<0.
\]
In addition, $\phi''(x)=\gamma_1(\gamma_1-1)C_1x^{\gamma_1-2}>0$.
In view of this, $\phi'(x)$ is increasing. Therefore, $\phi'(x)<0$ on
$(x^*,x_0^*)$, which implies $\phi(x)$ is decreasing. 
So $\phi(x)\leq\phi(x^*)\leq0$ on $(x^*,x_0^*)$.

Furthermore, using (\ref{cont-conds-case-II}), we can rewrite
(\ref{2nd-ineq}) as
\[
\frac{x^*-K}{\kappa_2}\leq \frac{(\rho+\la_2)(x^*-K)-x^* f_2}{\la_2},
\]
which in turn gives
\[
(\la_2-\kappa_2(\rho+\la_2-f_2))x^*\leq (\la_2-\kappa_2(\rho+\la_2))K.
\]
This inequality is equivalent to 
\[
x^*\leq \frac{(\la_2-\kappa_2(\rho+\la_2))K}
{\la_2-\kappa_2(\rho+\la_2-f_2)},
\]
because $\la_2-\kappa_2(\rho+\la_2-f_2)>0$ as shown in 
\lemref{cond3-ineq} in Appendix.

In view of (\ref{case-II-x^*-v2}), $x^*$ has to be bounded above by 
\[
X_0:=\min\left\{x_0^*,\, \frac{K\beta_2}{\beta_2-1},\,
\frac{(\la_2-\kappa_2(\rho+\la_2))K}
{\la_2-\kappa_2(\rho+\la_2-f_2)}\right\}.
\]
In view of \lemref{K-beta2} (Appendix), we have
\[
x_0^*=\frac{\rho K}{\rho-f_1}<\frac{K\beta_2}{\beta_2-1}
\]
Therefore, an upper bound for $x^*$ 
\[
X_0=\min\left\{\frac{K\beta_2}{\beta_2-1},\,
\frac{(\la_2-\kappa_2(\rho+\la_2))K}{\la_2-\kappa_2(\rho+\la_2-f_2)}\right\}.
\]

To obtain $x^*$, we only need to solve the first two equations 
in (\ref{cont-conds-case-II}). Eliminating $A_2$, we obtain  
\beq{x^*-Case-II}
C_1(x^*)^{\gamma_1}+\phi_0(x^*)=\frac{x^*-K}{\kappa_2},
\eeq
on $[K,x_0^*]$.

Let $\phi^*(x)=C_1x^{\gamma_1}+\phi_0(x)-{(x-K)}/{\kappa_2}$, Then it is easy
to check that $\phi_0(K)$ is positive, so is $\phi^*(K)$. In addition, 
$\phi^*(x_0^*)=(x_0^*-K)(1-1/\kappa_2)<0$ because $0<\kappa_2<1$. 
Furthermore, it can be shown that $(\phi^*)''(x)>0$, 
which implies that $(\phi^*)'(x)$
increasing. Using (\ref{1-ineq-VI-2-Case-II}) to obtain $(\phi^*)'(x)<0$,
which implies $\phi^*(x)$ is decreasing.
Therefore, $\phi^*(x)$ has a unique zero $x^*$ on $[K,x_0^*]$.

Recall that 
\[
x_0^*=\frac{\rho K}{\rho-f_1}
\mbox{ and }
C_1=\frac{x_0^*-K-\phi_0(x_0^*)}{(x_0^*)^{\gamma_1}}.
\]
Let $x^*$ be the solution of (\ref{x^*-Case-II}) over $[K,x_0^*]$ and let
\[
A_2=\frac{x^*-K}{\kappa_2(x^*)^{\beta_2}}.
\]

We have proved the following results.

\begin{thm}\bdd\label{HJB-Case-II}
Assume $\rho> f_1$.
Then the functions $v(x,i)$, $i=1,2$, given in 
{\rm (\ref{value-fn-Case-II})}
are continuous on $(0,\infty)$.
Moreover, $v(x,1)$ is differentiable on $(0,\infty)-\{x^*,x_0^*\}$
and $v(x,2)$ is differentiable on $(0,\infty)-\{x^*\}$.
If $x^*\leq X_0$, then they satisfy the HJB equations {\rm (\ref{HJB})}.
%
\end{thm}

\begin{rem}\bdd
{\rm
Note that $X_0\in[K,x_0^*]$. Recall that $\phi^*(x)$ is decreasing on
$[K,x_0^*]$. A sufficient condition
for $x^*\leq X_0$ is $\phi^*(X_0)\leq0$.
}
\end{rem}

\section{Verification Theorems and Numerical Examples}

First, we give two verification theorems depending on 
$\rho\leq f_1$ and $\rho> f_1$.
We only prove \thmref{VerificationThm1}. The proof of 
\thmref{VerificationThm2} can be given in a similar way.

\begin{thm}\label{VerificationThm1}\bdd
Assume the conditions of \thmref{HJB-Case-I}. Then,
$v(x,i)=V(x,i)$, $i=1,2$. Moreover, let 
$D=(0,\infty)\times\{1\}\cup(0,x^*)\times\{2\}$
denote the continuation region. Then
\[
\tau^*=\inf\{t\geq0;\, (S_t,\al_t)\not\in D\}
\]
is an optimal selling time.
\end{thm}

\nd{\it Proof.}
We only sketch the proof because it is similar to that of  
Zhang and Zhang \cite[Theorem 5]{ZhangZ}.
For any given stoppting time $\tau$ and $n=1,2,\ldots$, we have
\beq{verthm1ineq}
\begin{array}{rl}
v(x.i)\geq& Ee^{-\rho(\tau\wedge n)}v(S_{\tau\wedge n},\al_{\tau\wedge n})\\
=&Ee^{-\rho\tau}v(S_{\tau},\al_{\tau})I_{\{\tau<n\}}
+Ee^{-\rho n}v(S_{n},\al_{n})I_{\{\tau\geq n\}}\\
\geq &Ee^{-\rho\tau}(S_\tau-K)I_{\{\tau<n\}}
+Ee^{-\rho n}v(S_{n},\al_{n})I_{\{\tau\geq n\}}.\\
\end{array}
\eeq
Recall the linear growth of $v(x,i)$ and $Ee^{-\rho t}S_t\to0$ under (A2)
as $t\to\infty$. The second term goes to 0 and $n\to\infty$ 
Note also that the first term converges to 
$Ee^{-\rho\tau}(S_{\tau}-K)I_{\{\tau<\infty\}}=J(x,i,\tau)$.
It follows that $v(x,i)\geq J(x,i,\tau)$.

To show the equality, first if $(S_0,\al_0)=(x,i)\not\in D$, then $\tau^*=0$. 
This implies $v(x,i)=x-K=J(x,i,\tau^*)$.
If $(x,i)\in D$, then similarly as in (\ref{verthm1ineq}), we have
\[
\begin{array}{rl}
v(x.i)=& Ee^{-\rho(\tau^*\wedge n)}v(S_{\tau^*\wedge n},\al_{\tau^*\wedge n})\\
=&Ee^{-\rho\tau^*}v(S_{\tau^*},\al_{\tau^*})I_{\{\tau^*<n\}}
+Ee^{-\rho n}v(S_{n},\al_{n})I_{\{\tau^*\geq n\}}\\
=&Ee^{-\rho\tau^*}(S_{\tau^*}-K)I_{\{\tau^*<n\}}
+Ee^{-\rho n}v(S_{n},\al_{n})I_{\{\tau^*\geq n\}}\\
\to&Ee^{-\rho\tau^*}(S_{\tau^*}-K)I_{\{\tau^*<\infty\}}+0\\
=&J(x,i,\tau^*),
\end{array}
\]
as $n\to\infty$.
\hfill$\Box$

\begin{thm}\label{VerificationThm2}\bdd
Assume the conditions of \thmref{HJB-Case-II}. Then,
$v(x,i)=V(x,i)$, $i=1,2$. Moreover, let 
$D=(0,x_0^*)\times\{1\}\cup(0,x^*)\times\{2\}$
denote the continuation region. Then
\[
\tau^*=\inf\{t\geq0;\, (S_t,\al_t)\not\in D\}
\]
is an optimal selling time.
\end{thm}

\begin{cor}\bdd
Let ${\cal T}$ denote the class of almost sure finite
${\cal F}_t$ stopping times. Then,
\[
\sup_{\tau\in{\cal T}} E\(e^{-\rho\tau}\exp\int_0^\tau f(\al_t)dt\)<\infty.
\]
\end{cor}

\nd{\it Proof.}
Given $\al_0=i$, we have
\[
x E\(e^{-\rho\tau}\exp\int_0^\tau f(\al_t)dt\)I_{\{\tau<\infty\}}
\leq J(x,i,\tau)+K\leq V(x,i)+K\leq \max_i V(x,i)+K.
\]
Set $x=1$ to obtain
\[
\sup_{\tau\in{\cal T}} E\(e^{-\rho\tau}\exp\int_0^\tau f(\al_t)dt\)
\leq \max_i V(1,i)+K<\infty.\quad\Box
\]

\subsubsection*{Example 1 (Convergence to a Brownian motion).}

In this example, given $\e>0$, we consider 
\[
f_1=\mu+\frac{\sigma}{\sqrt{\e}},\
f_2=\mu-\frac{\sigma}{\sqrt{\e}},\
\la_1=\la_2=\frac{1}{\e}.\
\]

Using the asymptotic normality given in Yin and Zhang \cite[Theorem 5.9]{YinZ},
we can show that $S_t=S_t^\e$ converges weakly to 
\[
S_t^0=S_0e^{\mu t+\sigma W_t},\mbox{ as }\e\to0,
\]
where $W_t$ is a standard Brownian motion.
Such a limit is the solution to the stochastic 
differential equation
\[
\frac{dS_t}{S_t}=\(\mu+\frac{\sigma^2}{2}\)dt+\sigma dW_t.
\]

It is elementary to show that, as $\e\to0$,
\[
\beta_2
=\beta_2^\e\to \beta_0=\frac{-\mu+\sqrt{\mu^2+2\rho \sigma^2}}{\sigma^2}.
\]
This implies that $x^*=x^{*,\e}$ defined in (\ref{x*-alternative})
converges to the selling threshold $x^0=K\beta_0/(\beta_0-1)$ obtained in
${\O}$ksendal \cite[Example 10.2.2]{Oksendal}.

\begin{center}
\begin{figure}[htb]
\begin{center}
\mbox{\subfigure[{\small $\e=0.1$}]{
         \psfig{figure=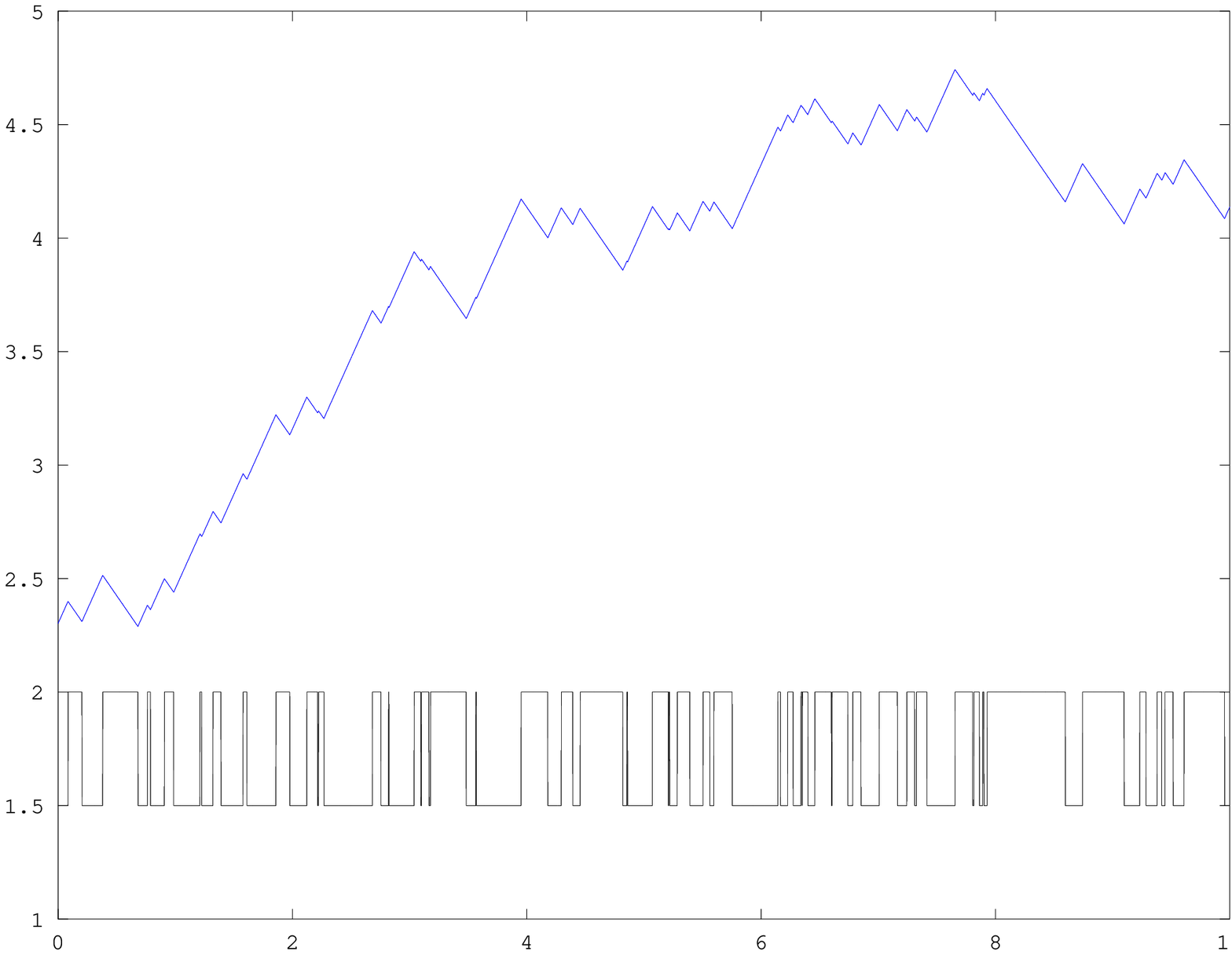,width=0.3\linewidth }} \quad
      \subfigure[{\small $\e=0.01$}]{
         \psfig{figure=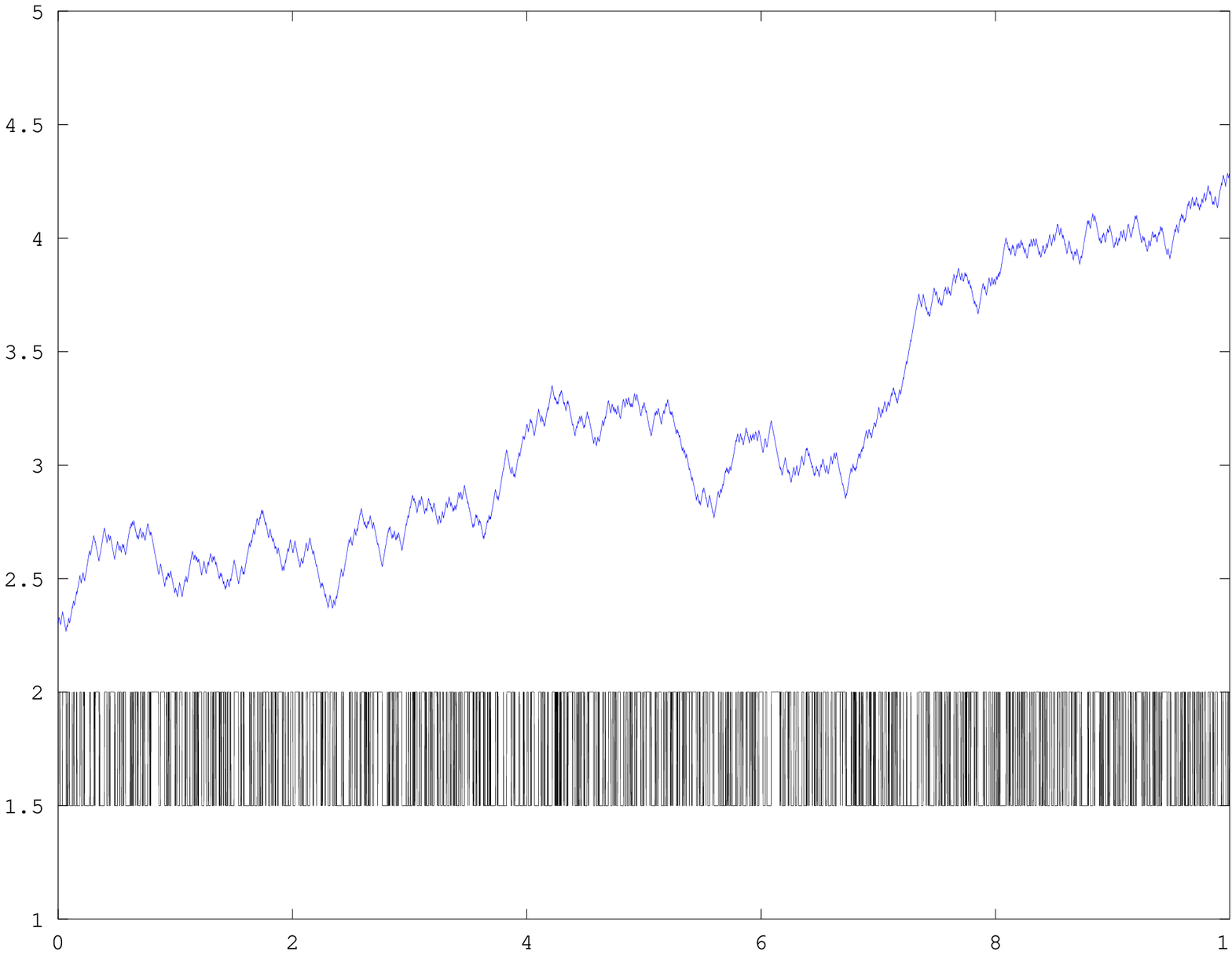,width=0.3\linewidth }} \quad
      \subfigure[{\small $\e=0.001$}]{
         \psfig{figure=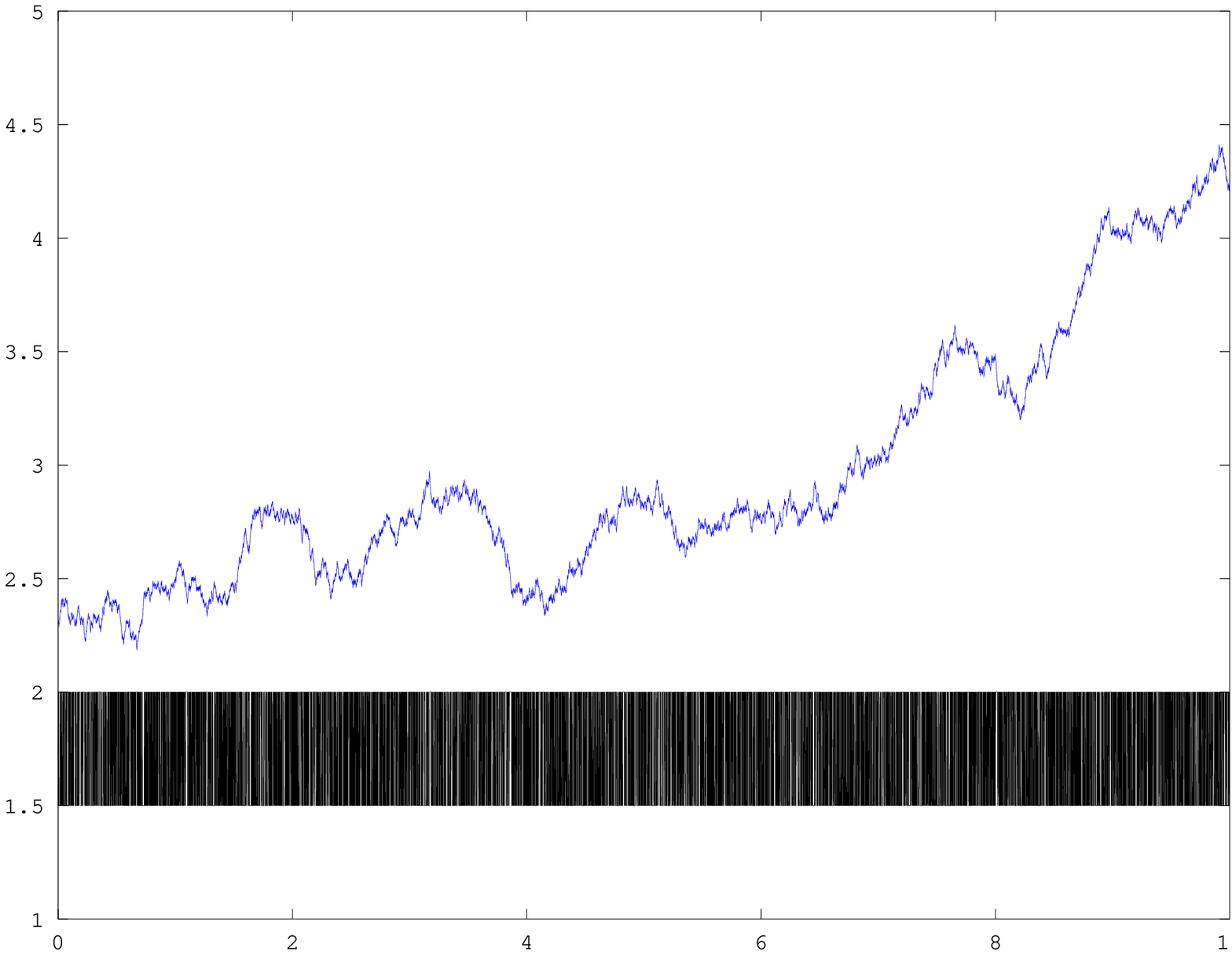,width=0.3\linewidth }}}
\end{center}
\caption{Monte Carlo Sample Paths: $(\log(S_t^\e),\al_t^\e)$}\label{MonteCarlo}
\end{figure}
\end{center}

Taking $\mu=0.2$ and $\sigma=0.3$,
we give sample paths of $\log(S_t^\e)$ and
$\al_t^\e$ with varying $\e$
in Figure~\ref{MonteCarlo}. It is clear from the pictures 
that as $\e$ gets smaller and smaller, the fluctuation of
$\al_t$ is more and more rapidly and the corresponding $S_t^\e$ approaches
to a GBM.

\subsubsection*{Example 2 (Case II).}

\begin{center}
\begin{figure}
\begin{center}
\epsfysize=2.0in
\centerline{\epsfbox{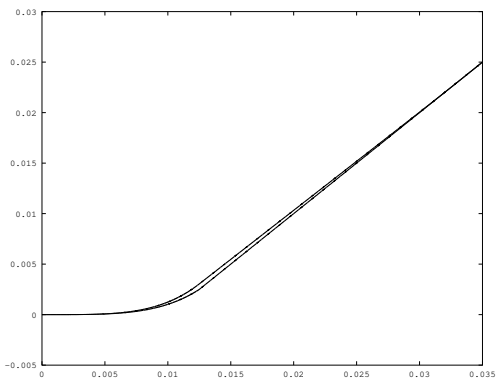}}
\vspace{-3ex}
\nd\caption{{\small Value Functions $v(x,1)$ and $v(x,2)$}}\label{VF-CaseII}
\end{center}
\end{figure}
\end{center}

In this example, we consider Case II with $\rho>f_1$ and
use the following parameters
\[
f_1=0.07,\,
f_2=-0.03,\,
\la_1=\la_2=1,\,
\rho=0.10,\,
K=0.01.
\]
Solving the equation (\ref{x^*-Case-II}) with $x_0^*=\rho K/(\rho-f_1)$, we
have $(x^*,x_0^*)=(0.012478,0.033333)$ and $X_0=0.013326$.
The corresponding value functions
are given in Figure~\ref{VF-CaseII}, in which 
$V(x,1)$ is given by the upper curve and $V(x,2)$ the lower one.

\subsubsection*{Example 3 (Model Calibration and a Market Test).}
First we give a model calibration method. We consider 
\[
f_1=\mu+\sigma_1\mbox{ and }
f_2=\mu+\sigma_2,
\]
with $\nu_1\sigma_1-\nu_2\sigma_2=0$.
Given $T$,
let $Y_t=\log(S_t/S_0)=\int_0^t f(\al_s)ds$, $0\leq t\leq T$.
Then, $\mu$ can be approximated by $Y_T/T$.
To estimate $\sigma_1$ and $\sigma_2$, given step size $\delta>0$, let 
$n\delta=T$,
\[
\Delta Z_k=\log(S_{(k+1)\delta})-\log(S_{k\delta}),
\mbox{ for } k=0,1,2,\ldots,n,
\]
and
$\overline Z=\(\sum_{k=0}^{n-1} \Delta Z_k\)/n$.
Then, $\overline Z\approx \delta \mu$.
In addition, using Yin an Zhang \cite[Theorem 5.9]{YinZ},
we can show
\[
E\(\Delta Z_k-\overline Z\)^2\approx \delta
\(\frac{2\la_1\la_2(\sigma_1+\sigma_2)^2}{(\la_1+\la_2)^3}\).
\]
Let 
\[
\sigma_0^2=\frac{\sum_{k=0}^{n-1} (\Delta Z_k-\overline Z)^2}{n-1}.
\]
Then, by the Law of Large Numbers, we have 
\[
\sigma_0^2\approx \delta\(\frac{2\la_1\la_2(\sigma_1+\sigma_2)^2}
{(\la_1+\la_2)^3}\).
\]
Using $\nu_1\sigma_1=\nu_2\sigma_2$, we have
\[
\sigma_1
=\frac{\sigma_0}{\sqrt{\delta}}\sqrt{\frac{\la_1(\la_1+\la_2)}{2\la_2}}
\mbox{ and }
\sigma_2
=\frac{\sigma_0}{\sqrt{\delta}}\sqrt{\frac{\la_2(\la_1+\la_2)}{2\la_1}},
\]
Finally, we estimate $\la_1$ and $\la_2$.
Let $R=\nu_1/\nu_2$. Then, $\la_2=R\la_1$. 
\[
\begin{array}{l}
R_1=\#\{k:\, \Delta Z_k<0\mbox{ and }\Delta Z_{k+1}\geq0\},\\
R_2=\#\{k:\, \Delta Z_k>0\mbox{ and }\Delta Z_{k+1}\leq0\}.\\
\end{array}
\]
Then, it follows that
\[
\frac{R_1}{\la_1}+\frac{R_2}{\la_2}=T.
\]
Therefore, the jump rates are given by
\[
\left\{\begin{array}{l}
\disp
\la_1=\frac{1}{T}\(R_1+\frac{R_2}{R}\),\\
\disp
\la_2=\frac{1}{T}\(R R_1+R_2\).\\
\end{array}\right.
\]

\begin{center}
\begin{figure}[htb]
\begin{center}
\mbox{\subfigure[{\small AAPL (2009/1/2--2013/3/28)}]{
         \psfig{figure=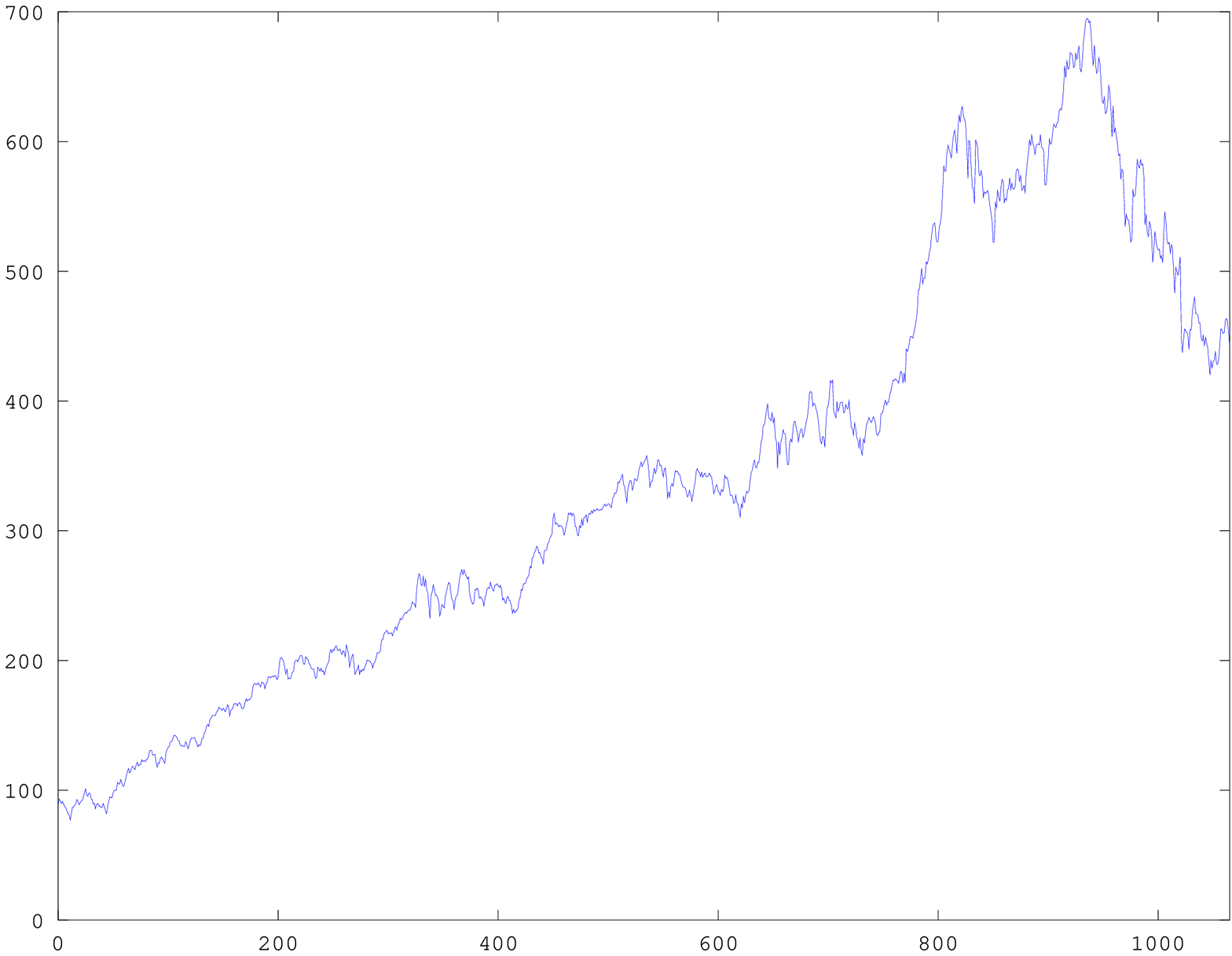,width=0.4\linewidth }} \quad
      \subfigure[{\small Value Functions}]{
         \psfig{figure=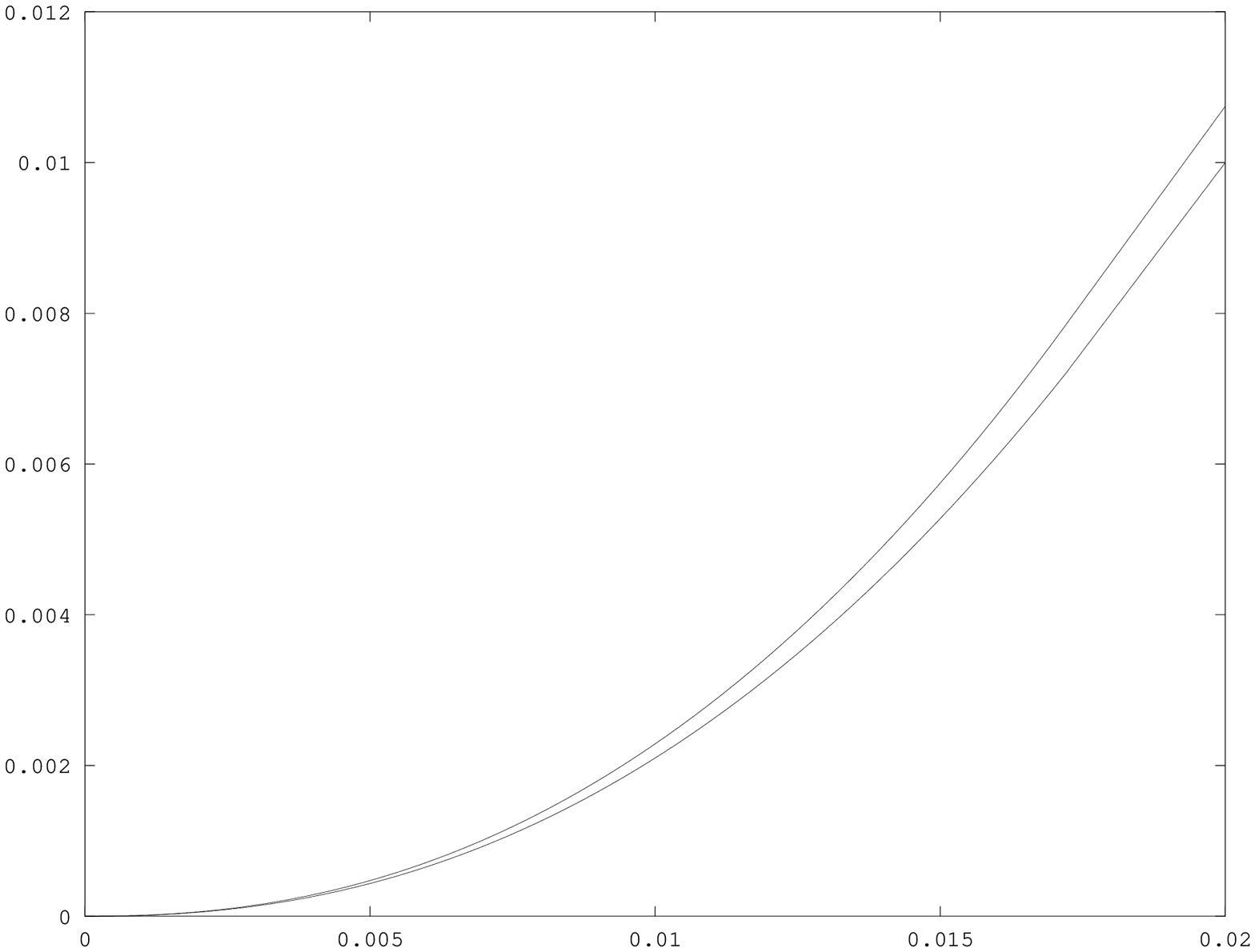,width=0.4\linewidth }}}
\end{center}
\caption{Apple Inc.  Daily Closes and Value Functions $v(x,1)$ and $v(x,2)$}\label{AAPL}
\end{figure}
\end{center}

We test our selling rules using Apple Inc. (AAPL) daily closing prices
during 2009/1/2 and 2013/3/28, sse Figure~\ref{AAPL} (a).
Suppose we owned 100 AAPL shares at the beginning of 2009.
We evaluate at the end of each half year during this period
based on that half year stock prices to determine if we should sell the shares
in the near future.

We assume the risk free rate to be $\rho=0.03$ and transaction cost 
$K=0.01$. We use the calibration method discussed earlier and obtain
the following results.

\begin{center}
{\small
\begin{tabular}{|c|c|c|c|c|c|}\hline
Periods&$f_1$&$f_2$&$\la_1$&$\la_2$&$\Phi(0.03)$\\ \hline 
1st half of 2009 &10.45&-10.61&100.48&124.23&-336.06\\ \hline
2nd half of 2009 &3.21&-2.32&102.15&141.44&-217.41\\ \hline
1st half of 2010 &3.06&-3.15&97.98&127.02&-83.18\\ \hline
2nd half of 2010 &2.27&-1.92&103.57&134.25 &-103.09 \\ \hline
1st half of 2011 &1.80&-1.85&117.19&125.00&-5.02 \\ \hline
2nd half of 2011 &3.01&-2.72&97.98 &107.95&-60.56 \\ \hline
1st half of 2012 &5.39 &-4.80& 108.21&127.19 &-185.32 \\ \hline
2nd half of 2012&4.89 &-5.13 &135.25 &130.95 &35.79 \\ \hline
\end{tabular}

\vspace*{0.1in}
\centerline{Table 1. Parameter Values at the End of Each Period}
}
\end{center}

Note that in all periods $\rho\leq f_1$. Therefore, only Case I applies
in this example.
In Table 1, we should hold through the next half year
if $\Phi(0.03)\leq 0$ and sell (following our 
selling rule) if $\Phi(0.03)>0$. Clearly, a selling decision has to be
made at the end of 2012. Using the parameter values 
$f_1=4.89$, $f_2=-5.13$, $\la_1=135.25$, and $\la_2=130.95$,
we obtain $x^*=0.017213$ and the corresponding value functions
$v(x,i)=V(x,i)$, $i=1,2$, which are plotted in Figure~\ref{AAPL} (b). 
Therefore, one should sell as soon as 
$\al_t$ turns to $2$ after the new year. This occurs on 
the second trading day (January 3) of 2013.
The shares should be sold at the close of that day at \$542.10/share.
As can be seen in this example, our selling rule helps to achieve 
the goal of letting your profits run and cutting your losses short.

\section{Conclusion}

In this paper, we considered an optimal stock selling rule under
a Markov chain model. The model is natural for financial markets 
due to its simple structure and the solutions obtained are 
intuitive and easy to implement.

It would be interesting to consider more general models with
multi-scale structure as treated in Yin and Zhang \cite{YinZ}
so as to capture both long-term and short-term market movements.
Such extension and related optimization problems could be 
subjects of future studies.


\section{Appendix}

\begin{lem}\label{beta2>1}\bdd
Under the assumption $\Phi(\rho)>0$, 
the bigger root of {\rm (\ref{char-eqn})} $\beta_2>1$.
\end{lem}

\nd{\it Proof.}
Recall the definition of $D_1$ and $D_2$ given in (\ref{D1-D2}).
It is easy to check $\Phi(\rho)>0$ implies
\[
D_2>D_1-f_1f_2.
\]
This leads to 
\[
\sqrt{D_1-4f_1f_2D_2}>D_1-2f_1f_2.
\]
Therefore, we have $\beta_2>1$. \hfill$\Box$

\begin{lem}\label{kappa2}\bdd
Let $\kappa_2={(\rho+\la_1-f_1\beta_2)}/{\la_1}$,
where $\beta_2$ is given in {\rm (\ref{char-root})}.
Then, $0<\kappa_2<1$.
\end{lem}

\nd{\it Proof.}
To see $\kappa_2>0$, it suffices to show
$\rho+\la_1>f_1\beta_2$. Recall that $f_2<0$. We only need to 
show $2(\rho+\la_1)f_2<D_1-\sqrt{D_1^2-4f_1f_2D_2}$,
with $D_1$ and $D_2$ given in (\ref{D1-D2}).

This is equivalent to 
\beq{kappa2-1}
\sqrt{D_1^2-4f_1f_2D_2}<D_1-2(\rho+\la_1)f_2.
\eeq
It is easy to check $D_1-2(\rho+\la_1)f_2=(\rho+\la_2)f_1-(\rho+\la_1)f_2>0$.
Square both sides of (\ref{kappa2-1}) to obtain
\[
D_1^2-4f_1f_2D_2<D_1^2-4(\rho+\la_1)f_2D_1+
4((\rho+\la_1)f_2)^2.
\]
Simplify this inequality to obtain
\[
D_2<(\rho+\la_1)(\rho+\la_2).
\]
This clearly holds. Therefore, $\kappa_2>0$.

Similarly, to show $\kappa_2<1$, it suffices to show
$\rho< f_1\beta_2$. This is equivalent to
\[
\sqrt{D_1^2-4f_1f_2D_2}>D_1-2\rho f_2.
\]
Square and simplify to obtain
$\rho \la_1 f_1>\rho\la_1 f_2$.
This holds because $f_1>0$ and $f_2<0$. \hfill$\Box$

\begin{lem}\label{K-beta2}\bdd
Under $\rho>f_1$, we have
\[
\frac{K\beta_2}{\beta_2-1}<\frac{\rho K}{\rho-f_1}.
\]
\end{lem}

\nd{\it Proof.}
It is easy to see this inequality is equivalent to $f_1\beta_2>\rho$.
Using the definition of $\beta_2$ and noting that $f_2<0$, we have
\[
\sqrt{D_1^2-4f_1f_2D_2}>D_1-2\rho f_2.
\]
If $D_1-2\rho f_2<0$, then we are done. Otherwise, square both sides to
obtain
\[
D_1^2-4f_1f_2D_2>D_1^2-4\rho f_2 D_1+4\rho^2f_2^2.
\]
Simplify this inequality to have 
\[
\rho\la_1f_1>\rho\la_1 f_2,
\]
which clearly holds because $f_1>0$ and $f_2<0$.

\begin{lem}\label{cond3-ineq}\bdd
Under $\rho>f_1$, we have
\beq{app-cond3}
\la_2-\kappa_2(\rho+\la_2-f_2)>0,\\
\eeq
which implies $\la_2-\kappa_2(\rho+\la_2)>0$.
\end{lem}

\nd{\it Proof.}
Let $H_0=\rho+\la_2-f_2>0$.
Using the definition of $\kappa_2$, (\ref{app-cond3}) is 
equivalent to 
\[
H_0(\rho+\la_1-f_1\beta_2)<\la_1\la_2.
\]
Therefore, we have
\[
(\rho+\la_1)H_0-\(\frac{D_1-\sqrt{D_1^2-4f_1f_2D_2}}{2f_2}\)H_0<\la_1\la_2.
\]
Multiply both sides by $(2f_2)$ and rearrange the terms to obtain
\[
H_0\sqrt{D_1^2-4f_1f_2D_2}>H_0D_1-2f_2[(\rho+f_1)H_0-\la_1\la_2].
\]
Square both sides and simplify to have
\[
(\rho+\la_1-f_1)H_0-\la_1\la_2>0,
\]
which is exactly the assumption $\phi(\rho)>0$. 
This completes the proof. \hfill$\Box$

\end{document}